\documentclass[twocolumn, pra, showpacs,superscriptaddress]{revtex4}
\usepackage{amsfonts}
\usepackage{amsmath}
\usepackage{amssymb}
\usepackage{graphicx}

\begin{document}

\title{Atomic number fluctuations in a mixture of two spinor condensates}
\author{Jie Zhang}
\affiliation{Institute of Theoretical Physics, Shanxi University, Taiyuan 030006,
People's Republic of China}
\author{Z. F. Xu}
\affiliation{Department of Physics, Tsinghua University, Beijing 100084, People's
Republic of China}
\author{L. You}
\affiliation{Department of Physics, Tsinghua University, Beijing 100084, People's
Republic of China}
\author{Yunbo Zhang}
\email{ybzhang@sxu.edu.cn}
\affiliation{Institute of Theoretical Physics, Shanxi University, Taiyuan 030006,
People's Republic of China}

\begin{abstract}
We study particle number fluctuations in the quantum ground states of a
mixture of two spin-1 atomic condensates when the interspecies spin-exchange
coupling interaction $c_{12}\beta$ is adjusted. The two spin-1 condensates
forming the mixture are respectively ferromagnetic and polar in the absence
of an external magnetic (B-) field. We categorize all possible ground states
using the angular momentum algebra and compute their characteristic atom
number fluctuations, focusing especially on the the AA phase (when $%
c_{12}\beta >0$), where the ground state becomes fragmented and atomic
number fluctuations exhibit drastically different features from a single
stand alone spin-1 polar condensate. Our results are further supported by
numerical simulations of the full quantum many-body system.
\end{abstract}

\pacs{03.75.Mn, 67.60.Bc, 67.85.Fg}
\maketitle

%\affiliation{$^{3}$School of Physics, Georgia
%Institute of Technology, Atlanta, Georgia 30332, USA}

\section{Introduction}

Since the first production of an atomic $^{23}$Na condensate in an optical
trap \cite{Stamper-Kurn}, spin degrees of freedom for condensed atoms become
accessible, which has since given rise to a rich variety of phenomena such
as domain formations \cite{Stenger}, spin mixing dynamics \cite{MSChang},
topological defects \cite{Topology}, etc. The properties of a
three-component ($F=1$) spinor condensate are first studied by Ho \cite{Ho}
and Ohmi \cite{Ohmi}. Many predictions are experimentally verified \cite%
{Stenger}, the most fundamental property concerns the existence of two
different phases: the so-called polar and ferromagnetic states, respectively
corresponding to the $F=1$ state of $^{23}$Na and $^{87}$Rb atomic
condensates.

In contrast to a scalar condensate, both spatial and internal spin part of
the wave functions are required to discuss a spinor condensate. For both $F=1
$ state of $^{23}$Na and $^{87}$Rb atoms, density-density interactions are
significantly larger than the spin-exchange interactions. The single
spatial-mode approximation (SMA) \cite{Ho,Ohmi,Law,Pu,Yi} is often adopted,
whereby one adopts mean field approximation (MFT) to determine the
condensate spatial wave function neglecting their spin dependence. The spin
degrees of freedom is then considered assuming the spatial wave function is
identical. The spin related properties can be investigated use either a
\textquotedblleft mean-field\textquotedblright\ theory or \textquotedblleft
semiclassical treatment\textquotedblright \cite{Ho,Ohmi}, where the spinor
is described by a vector formed by three $c$ numbers $\zeta _{\alpha }(\zeta
_{\alpha }^{\ast }) (\alpha =1,0,-1)$, or use many body theory \cite{Law,Pu}
treating the spinor from the bosonic mode as operators $\hat{a}_{\alpha },$
satisfying $\left[ \hat{a}_{\alpha },\hat{a}_{\beta }^{\dag }\right] =\delta
_{\alpha \beta }$. The quantum ground states for a spin-1 condensate have
been studied extensively.

It was initially predicted that the ground state of $^{23}$Na BEC ($c_{2}>0$%
) is either polar ($n_{0}=N$) or anti-ferromagnetic ($n_{1}=n_{-1}=N/2)$ in
the mean-field theory \cite{Ho,Ohmi}. A quantum treatment based on the SMA
by Law, Pu and Bigelow \cite{Law,Pu} revealed, however, the ground state for
$^{23}$Na atoms is actually a spin singlet with properties ($%
n_{1}=n_{0}=n_{-1}=N/3 $), drastically different from polar state predicted
within the mean field theory. Further studies pointed out that this spin
singlet state is a fragmented condensate with anomalously large number
fluctuations and thus has fragile stability \cite{HoYip,Mueller06}. The
remarkable nature of this fragmentation is characterized by three
macroscopic eigenvalues (see above) for the single particle reduced density
matrix, which is capable of exhibiting anomalously large atom number
fluctuations $\Delta n_{1,0,-1}\sim N$.

The interests in spinor condensates extend to higher spins \cite%
{Ciobanu,Spin3} and quantum ground states are already well known and
categorized for spin-2 condensates \cite{Ueda}. Atomic Feshbach resonance
was implemented in a double condensate, enabling tunable interactions, whose
effects on superfluid dynamics and controlled phase separations are observed
\cite{MixtureE}. The spin exchange interaction between individual atoms can
be precisely tuned through \textquotedblleft optical Feshbach
resonances\textquotedblright\ \cite{OFR} by adjusting the two s-wave
scattering lengths $a_{0}$ and $a_{2}$. This inspired several recent
theoretical studies on mixtures of spinor condensates \cite%
{Xu09,Xu2,Shi06,Shi09} and tunable or controlled dynamics \cite{W. Zhang}.

In this paper we report anomalous fluctuations for the numbers of atoms in a
binary mixture of spin-1 condensates. We hope to stimulate experiments,
using the most relevant experimental case, the mixture of $^{23}$Na (polar)
and $^{87}$Rb (ferromagnetic) condensates in their $F=1$ manifold, as an
example. The quantum spin properties in the special ground state of the AA
phase, where the interspecies anti-ferromagnetic spin-exchange is large
enough to polarize both species but forming an maximally entangled state
between two species, are studied and we give the exact number fluctuations
distribution. Then we resort to numerical diagonalizations to show that
particle numbers and number fluctuations undergo dramatic changes as
inter-species coupling $c_{12}\beta $ varies.

\section{The model Hamiltonian for the mixture}

Intra-condensate atomic interaction takes the form $V_{j}(\mathbf{r}%
)=(\alpha _{j}+\beta _{j}\mathbf{F}_{j}\cdot \mathbf{F}_{j})\delta (\mathbf{r%
})$ with $j=1,2$ for the ferromagnetic ($\beta _{1}<0$) and polar ($\beta
_{2}>0$) atoms. The inter-species interaction between the ferromagnetic and
polar atoms is described as $V_{12}(\mathbf{r})=\frac{1}{2}(\alpha +\beta
\mathbf{F}_{1}\cdot \mathbf{F}_{2}+\gamma P_{0})\delta (\mathbf{r})$ \cite%
{Xu09}, which is more complicated because collision can occur in the total
spin $F_{\mathrm{tot}}=1$ channel between different atoms, in contrast to
intra-condensate interactions between identical atoms \cite{Luo,Xu09}. The
parameters $\alpha ,\beta $, and $\gamma $ are related to the $s$-wave
scattering lengths in the the total spin channels \cite{Xu09}, analogous to
spin-1 condensates \cite{Ho,Ohmi}. $P_{0}$ projects an inter-species pair
into spin singlet state and $\mu =M_{1}M_{2}/(M_{1}+M_{2})$ denotes the
reduced mass for the pair of atoms, one each from the two different species
with masses $M_{1}$ and $M_{2}$ respectively.

Our model Hamiltonian is given by
\begin{eqnarray}
\hat{H} &=&\hat{H}_{1}+\hat{H}_{2}+\hat{H}_{12}, \\
\hat{H}_{1} &=&\int d\mathbf{r}\left\{ \hat{\Psi}_{i}^{\dag }(\frac{\hbar
^{2}}{2M_{1}}\nabla ^{2}+U_{1})\hat{\Psi}_{i}+\frac{\alpha _{1}}{2}\hat{\Psi}%
_{i}^{\dag }\hat{\Psi}_{j}^{\dag }\hat{\Psi}_{j}\hat{\Psi}_{i}\right.
\notag \\
&&\left. +\frac{\beta _{1}}{2}\hat{\Psi}_{i}^{\dag }\hat{\Psi}_{j}^{\dag }%
\mathbf{F}_{1il}\cdot \mathbf{F}_{1jk}\hat{\Psi}_{k}\hat{\Psi}_{l}\right\} ,
\notag \\
\hat{H}_{12} &=&\frac{1}{2}\int d\mathbf{r}\left\{ \alpha \hat{\Psi}%
_{i}^{\dag }\hat{\Phi}_{j}^{\dag }\hat{\Phi}_{j}\hat{\Psi}_{i}\right.
\notag \\
&&\left. +\beta \hat{\Psi}_{i}^{\dag }\hat{\Phi}_{j}^{\dag }\mathbf{F}%
_{1il}\cdot \mathbf{F}_{2jk}\hat{\Phi}_{k}\hat{\Psi}_{l}+\frac{\gamma }{3}%
\hat{O}^{\dag }\hat{O}\right\} .
\end{eqnarray}%
$H_{2}$ is identical to $H_{1}$ except for the substitution of subscript $1$
by $2$ and $\hat{\Psi}_{i}$ by $\hat{\Phi}_{i}$. The latter two are atomic
field operators for the spin state $\left\vert 1,i\right\rangle $. $\hat{O}=%
\hat{\Psi}_{1}\hat{\Phi}_{-1}-\hat{\Psi}_{0}\hat{\Phi}_{0}+\hat{\Psi}_{-1}%
\hat{\Phi}_{1}$.

\begin{figure}[tbp]
\includegraphics[width=3.0in]{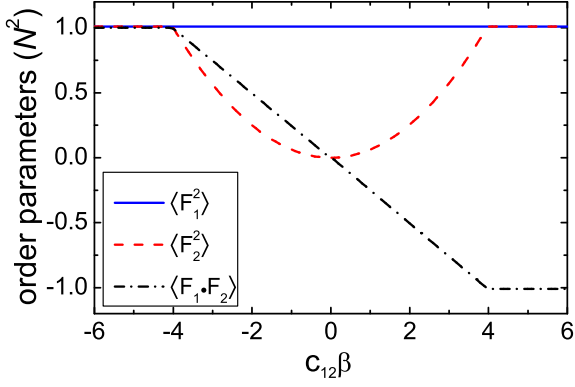}
\caption{(Color online) The dependence of ground-state phases on $\protect%
\beta $ at fixed values of $c_{1}\protect\beta _{1}=-1,$ $c_{2}\protect\beta %
_{2}=2$. All interaction parameters are in units of $|c_{1}\protect\beta %
_{1}|$. }
\label{fig1}
\end{figure}

We adopt the SMA \cite{Law,Pu,Yi} for each of the two spinor condensates
with modes $\Psi (\mathbf{r})$ and $\Phi (\mathbf{r})$, i.e., setting
\begin{equation*}
\hat{\Psi}_{i}=\hat{a}_{i}\Psi ,\text{\qquad }\hat{\Phi}_{i}=\hat{b}_{i}\Phi
,
\end{equation*}%
with $\hat{a}_{i}$ ($\hat{b}_{i}$) the annihilation operator for the
ferromagnetic (polar) atoms satisfying $\left[ \hat{a}_{i},\hat{a}_{j}\right]
=0$ and $\left[ \hat{a}_{i},\hat{a}_{j}^{\dag }\right] =\delta _{ij}$ (and
the same form of commutations for $\hat{b}_{i}$). The spin-dependent
Hamiltonian for our mixture mode then reads
\begin{eqnarray}
\hat{H} &=&\frac{c_{1}\beta _{1}}{2}\mathbf{\hat{F}}_{1}^{2}+\frac{%
c_{2}\beta _{2}}{2}\mathbf{\hat{F}}_{2}^{2}  \notag \\
&&+\frac{c_{12}\beta }{2}\mathbf{\hat{F}}_{1}\cdot \mathbf{\hat{F}}_{2}+%
\frac{c_{12}\gamma }{6}\hat{\Theta}_{12}^{\dag }\hat{\Theta}_{12},
\label{Ham}
\end{eqnarray}%
where $\mathbf{\hat{F}}_{1}=\hat{a}_{i}^{\dag }\mathbf{F}_{1ij}\hat{a}_{j}$ (%
$\mathbf{\hat{F}}_{2}=\hat{b}_{i}^{\dag }\mathbf{F}_{2ij}\hat{b}_{j}$) are
defined in terms of the $3\times 3$ spin-1 matrices $\mathbf{F}_{1ij}$ $(%
\mathbf{F}_{2ij})$, and
\begin{equation*}
\hat{\Theta}_{12}^{\dag }=\hat{a}_{0}^{\dag }\hat{b}_{0}^{\dag }-\hat{a}%
_{1}^{\dag }\hat{b}_{-1}^{\dag }-\hat{a}_{-1}^{\dag }\hat{b}_{1}^{\dag },
\end{equation*}%
creates a singlet pair with one atom each from the two species, similar to
the following two
\begin{equation*}
\hat{A}^{\dag }=(\hat{a}_{0}^{\dag })^{2}-2\hat{a}_{1}^{\dag }\hat{a}%
_{-1}^{\dag },\text{\qquad }\hat{B}^{\dag }=(\hat{b}_{0}^{\dag })^{2}-2\hat{b%
}_{1}^{\dag }\hat{b}_{-1}^{\dag },
\end{equation*}%
for intra-species spin-singlet pairs \cite{HoYip}. The interaction
coefficients are $c_{1}=\int d\mathbf{r}\left\vert \Psi (r)\right\vert ^{4},$
$c_{2}=\int d\mathbf{r}\left\vert \Phi (r)\right\vert ^{4}$ and $c_{12}=\int
d\mathbf{r}\left\vert \Psi (r)\right\vert ^{2}\left\vert \Phi (r)\right\vert
^{2},$ which can be tuned through the control of the trapping frequency.
Here we focus on the spin-dependent part when we assume that the two species
are sufficiently overlapped. The scattering properties between any pairs of 
atoms in specific Zeeman hyperfine
component states are determined from their corresponding interaction potentials
observing the symmetries of the two states. We notice that under the so-called 
degenerate internal-states approximation (DIA) the $\gamma $ term vanishes \cite{DIA}.
Within the DIA, the Zeeman hyperfine state of an alkali atom
is expanded in terms of the electronic (valence electron) and its corresponding nuclear spin states.
Between any two atoms, the corresponding interaction potential is constructed from
the appropriate weighted spin singlet/triplet potentials when expanded out correspondingly.
When the spin singlet and triplet potentials are further approximated by low energy
scattering pseudo-potentials, the total scattering properties are determined completely by
two scattering lengths of the singlet and triplet potentials. The parameter $\gamma$ is
therefore no longer needed and the Hamiltonian (\ref{Ham})
finally reduces to
\begin{equation}
\hat{H}_{A}=\frac{c_{1}\beta _{1}}{2}\mathbf{\hat{F}}_{1}^{2}+\frac{%
c_{2}\beta _{2}}{2}\mathbf{\hat{F}}_{2}^{2}+\frac{c_{12}\beta }{2}\mathbf{%
\hat{F}}_{1}\cdot \mathbf{\hat{F}}_{2}.  \label{Ham2}
\end{equation}

\section{The Ground States}

We rewrite the the Hamiltonian (\ref{Ham2}) as
\begin{equation}
\hat{H}=a\mathbf{\hat{F}}_{1}^{2}+b\mathbf{\hat{F}}_{2}^{2}+c\mathbf{\hat{F}}%
^{2},  \label{mH}
\end{equation}%
with $a=c_{1}\beta _{1}/2-c_{12}\beta /4$, $b=c_{2}\beta _{2}/2-c_{12}\beta
/4$, and $c=c_{12}\beta /4,$ $\mathbf{\hat{F}}=\mathbf{\hat{F}}_{1}+\mathbf{%
\hat{F}}_{2}$ is the total spin operator. The eigenstates of (\ref{mH}) are
the common eigenstates for the commuting operators $\mathbf{\hat{F}}_{1}^{2},%
\mathbf{\hat{F}}_{2}^{2},\mathbf{\hat{F}}^{2}$, and $\hat{F}_{z}$, given by
\begin{equation*}
\left\vert F_{1},F_{2},F,m\right\rangle
=\sum_{m_{1}m_{2}}C_{F_{1,}m_{1};F_{2,}m_{2}}^{F,m}\left\vert
F_{1},m_{1}\right\rangle \left\vert F_{2},m_{2}\right\rangle ,
\end{equation*}%
with the uncoupled basis states
\begin{equation*}
\left\vert F_{1},m_{1}\right\rangle =Z_{1}^{-\frac{1}{2}}(\hat{F}%
_{1-})^{F_{1}-m_{1}}(\hat{a}_{1}^{\dag })^{F_{1}}(\hat{A}^{\dag })^{\left(
N_{1}-F_{1}\right) /2}\left\vert 0\right\rangle ,\hskip12pt
\end{equation*}%
and analogously for $\left\vert F_{2},m_{2}\right\rangle $ which span a
Hilbert space of dimension $(N_{j}+1)(N_{j}+2)/2$ \cite{Ueda}. $C$ is the
Clebsch-Gordon coefficient, $Z_{j}$ is a normalization constant, and $\hat{F}%
_{j-}$ is the lowering operator for $m_{j}$. The corresponding eigenenergy
is
\begin{equation}
E=aF_{1}(F_{1}+1)+bF_{2}(F_{2}+1)+cF(F+1).  \label{Eval}
\end{equation}%
Given $N_{j}$, the allowed values of $F_{j}$ are $F_{j}=0,2,4,\cdots N_{j}$
if $N_{j}$ is even and $F_{j}=1,3,5,\cdots N_{j}$ if $N_{j}$ is odd,
satisfying $\left\vert F_{1}-F_{2}\right\vert \leqslant F\leqslant
F_{1}+F_{2}$.

We next consider the special case of $N_{1}=N_{2}=N$ and for $N$ even.
Minimizing the energy (\ref{Eval}), we can get the ground state phases
determined by different parameters $a$, $b$, and $c$, or equivalently $%
c_{1}\beta _{1},$ $c_{2}\beta _{2},$ and $c_{12}\beta $. Taking $N=100$ for
example, Figure \ref{fig1} presents the results for the order parameters $%
\langle \mathbf{\hat{F}}_{1}^{2}\rangle $, $\langle \mathbf{\hat{F}}%
_{2}^{2}\rangle $, and $\langle \mathbf{\hat{F}}_{1}\cdot \mathbf{\hat{F}}%
_{2}\rangle =(\langle \mathbf{\hat{F}}^{2}\rangle -\langle \mathbf{\hat{F}}%
_{1}^{2}\rangle -\langle \mathbf{\hat{F}}_{2}^{2}\rangle )/2$ \cite{Xu09} as
$c_{12}\beta $ changes but fixed $c_{1}\beta _{1}=-1$, $c_{2}\beta _{2}=2$
(in units of $|c_{1}\beta _{1}|$), which are found to agree very well with
the mean field ones obtained from simulated annealing \cite{Xu09}.

\begin{widetext}
\begin{center}
\begin{table}[tbh]
\begin{tabular}{|l|l|l|}
\hline $\text{Phases}$ & $\text{Parameter range}$ & $\text{Ground
states }\left\vert F_{1},F_{2},F,F\right\rangle $ \\ \hline
$FF$ & $-\infty <c_{12}\beta <\frac{-(2N-1)}{N}c_{2}\beta _{2}$ & $%
C_{N,N;N,N}^{2N,2N}\left\vert N,N\right\rangle \left\vert
N,N\right\rangle $
\\ \hline
$MM_{-}$ & $\frac{-(2N-1)}{N}c_{2}\beta _{2}<c_{12}\beta < 0$ & $%
C_{N,N;F_{2},F_{2}}^{N+F_{2},N+F_{2}}\left\vert N,N\right\rangle
\left\vert F_{2},F_{2}\right\rangle $ \\ \hline
$MM_{+}$ & $0<c_{12}\beta <\frac{2N-1}{N+1}c_{2}\beta _{2}$ & $\underset{%
m_{1},m_{2}}{\sum
}C_{N,m_{1};F_{2,}m_{2}}^{N-F_{2},N-F_{2}}\left\vert
N,m_{1}\right\rangle \left\vert F_{2},m_{2}\right\rangle $ \\ \hline
$AA$ & $\frac{2N-1}{N+1}c_{2}\beta _{2}<c_{12}\beta <+\infty $ & $\underset{%
m_{1}}{\sum }C_{N,m_{1};N,-m_{1}}^{0,0}\left\vert
N,m_{1}\right\rangle \left\vert N,-m_{1}\right\rangle $ \\ \hline
%\label{tab1}
\end{tabular}%
\caption{Quantum phases for the ground state.}
\label{TGS}
\end{table}
\end{center}
\end{widetext}Overall, there are four different phases in the general case,
separated by three critical points ${-(2N-1)}c_{2}\beta _{2}/{N}$, $0$, and $%
{(2N-1)}c_{2}\beta _{2}/({N+1})$ (corresponding to $-4$, $0$, and $4$ in
Fig. \ref{fig1}). The extreme states $\left\vert
F_{1},F_{2},F,F\right\rangle $ are classified into FF, MM$_{-}$, MM$_{+}$,
and AA phases as in Table 1. Other degenerate states are found by repeated
applications of $\hat{F}_{1-}+\hat{F}_{2-}$,
\begin{eqnarray}
\left\vert F_{1},F_{2},F,m\right\rangle =(\hat{F}_{1-}+\hat{F}%
_{2-})^{F-m}\left\vert F_{1},F_{2},F,F\right\rangle ,
\end{eqnarray}%
with $m=0,\pm 1,\cdots \pm F$.

The ground state for a spin-1 polar condensate is fragmented \cite{HoYip},
described by a spin singlet state of the form $(\hat{B}^{\dag
})^{N/2}\left\vert 0\right\rangle $. $\hat{B}^{\dag }$ ($\hat{B}$) is
invariant under rotations and commutes with $\hat{F}_{2}$ and $\hat{F}_{2z}$%
. For ferromagnetic condensate, the condensate ground state favors all atoms
aligned along the same direction, i.e., takes the form $(\hat{a}_{1}^{\dag
})^{N}\left\vert 0\right\rangle $, and is more stable.
%Without the external magnetic field,
%the ground state is degenerated, there are a lot of discontinuous directions
%with different magnetization $m_{1}$. We can enumerate them using lowering
%operator as $\left\vert F_{1}=N,m_{1}\right\rangle =(\hat{F}%
%_{1-})^{F_{1}-m_{1}}(\hat{a}_{1}^{\dag })^{N}\left\vert 0\right\rangle ,$
%with $m_{1}=0,\pm 1,\pm 2,...\pm N.$
When mixing the two together, we expect the polar atoms are strongly
affected, while the back action on to the more stable ferromagnetic atoms is
negligible.

The FF phase may be simply described as $Z^{1/2}(\hat{a}_{1}^{\dag })^{N}(%
\hat{b}_{1}^{\dag })^{N}\left\vert 0\right\rangle $, with all polar atoms
slaved into the same direction as the ferromagnetic ones. For $%
-(2N-1)c_{2}\beta _{2}/N<c_{12}\beta <0$, the MM$_{-}$ phase arises when
polar atoms are partly polarized in the same direction with the
ferromagnetic ones, as if there were a finite B-field. Increase of the
coupling interaction ($\left\vert c_{12}\beta \right\vert $) breaks singlet
pairs in polar atoms one by one with $\Delta F_{2}=2$, and gives rise to
stepwise increases of the total spin. Similar to the FF phase, the MM$_{-}$
phase has a simpler form, $Z^{1/2}(\hat{a}_{1}^{\dag })^{N}(\hat{b}%
_{1}^{\dag })^{F_{2}}(\hat{B}^{\dag })^{(N-F_{2})/2}\left\vert
0\right\rangle $, with its energy $E=c_{1}\beta _{1}N(N+1)/2+c_{2}\beta
_{2}F_{2}(F_{2}+1)/2+c_{12}\beta NF_{2}/2$. This phase corresponds to the
exact ground state for $F_{2}-1<-c_{12}\beta N/2c_{2}\beta _{2}-1/2<F_{2}+1$
\cite{Ueda}, and we find the order parameters%
\begin{eqnarray}
\langle \hat{F}_{1}^{2}\rangle \ &=&N(N+1),  \notag \\
\langle \hat{F}_{2}^{2}\rangle \ &=&\left( \frac{N}{2c_{2}\beta _{2}}\right)
^{2}\left( c_{12}\beta \right) ^{2}-\frac{1}{4},  \notag \\
\langle \hat{F}_{1}\cdot \hat{F}_{2}\rangle \ &=&-\frac{N^{2}}{2c_{2}\beta
_{2}}c_{12}\beta -\frac{N}{2},  \label{MM+}
\end{eqnarray}%
agree with the mean-field results \cite{Xu09} to terms $\sim 1/N$.

For $0<c_{12}\beta <({2N-1})c_{2}\beta _{2}/({N+1})$, however, the MM$_{+}$
phase favors polar atoms polarized opposite to the ferromagnetic atoms
resulting in a decreased total spin. This situation is more complicated
because all states satisfying $m_{1}+m_{2}=N-F_{2}$ are involved. The ground
state energy is $E=c_{1}\beta _{1}N(N+1)/2+c_{2}\beta
_{2}F_{2}(F_{2}+1)/2-c_{12}\beta (N+1)F_{2}/2$, and we find that
\begin{eqnarray}
\langle \hat{F}_{1}^{2}\rangle \ &=&N(N+1),  \notag \\
\text{\ }\langle \hat{F}_{2}^{2}\rangle \ &=&\left( \frac{N+1}{2c_{2}\beta
_{2}}\right) ^{2}\left( c_{12}\beta \right) ^{2}-\frac{1}{4},  \notag \\
\langle \hat{F}_{1}\cdot \hat{F}_{2}\rangle \ &=&-\frac{(N+1)^{2}}{%
2c_{2}\beta _{2}}c_{12}\beta -\frac{N+1}{2}.  \label{MM-}
\end{eqnarray}%
The stepwise fine structure of the order parameters is not included in Eqs. (%
\ref{MM+}) and (\ref{MM-}). When $c_{12}\beta >({2N-1})c_{2}\beta _{2}/({N+1}%
)$, again all states satisfying $m_{1}+m_{2}=0$ are included and the AA
phase is described by a singlet state \cite{Xu2},
\begin{eqnarray}
\left\vert N,N,0,0\right\rangle =\underset{m_{1}=-N}{\overset{N}{\sum }}%
C_{N,m_{1};N,-m_{1}}^{0,0}\left\vert N,m_{1}\right\rangle \left\vert
N,-m_{1}\right\rangle. \hskip 12pt
\end{eqnarray}%
We find interestingly that the total spin vanishes, while the species-spins
satisfy $\langle \hat{F}_{1}^{2}\rangle =\langle \hat{F}_{2}^{2}\rangle
=N(N+1)$ (see Fig. \ref{fig1}).

\section{Atomic number fluctuations}

As a special case, the AA phase is a singlet state, which enables us to
derive analytically the average particle numbers and their associated
fluctuations. With the help of the methods developed in a related cavity QED
problem \cite{Ying Wu}, we expand the eigenvectors\ $\left\vert
N,m_{1}\right\rangle $ in terms of the Fock states which are defined as $%
\hat{n}_{\alpha }^{(j)}\left\vert
n_{1}^{(j)},n_{0}^{(j)},n_{-1}^{(j)}\right\rangle =n_{\alpha
}^{(j)}\left\vert n_{1}^{(j)},n_{0}^{(j)},n_{-1}^{(j)}\right\rangle ,\alpha
=0,\pm 1$. Explicitly, for even and odd $m_{1}$, we find respectively
\begin{widetext}
\begin{eqnarray}
\left\vert N,m_{1}\right\rangle
&=&\sum_{k=0}^{(N-|m_{1}|)/2}B_{m_{1}k}^{(e)}\left\vert \frac{N+m_{1}}{2}%
-k,2k,\frac{N-m_{1}}{2}-k\right\rangle , \notag \\
\left\vert N,m_{1}\right\rangle
&=&\sum_{k=0}^{(N-|m_{1}|-1)/2}B_{m_{1}k}^{(o)}\left\vert \frac{N+m_{1}-1}{2}%
-k,2k+1,\frac{N-m_{1}-1}{2}-k\right\rangle,
\end{eqnarray}%
\end{widetext}where,%
\begin{eqnarray}
B_{m_{1}k}^{(e)} &=&\frac{(\sqrt{2})^{2k}(\frac{(N+m_{1})!(N-m_{1})!}{2N!}%
)^{1/2}\sqrt{N!}}{[(2k)!(\frac{N-m_{1}}{2}-k)!(\frac{N+m_{1}}{2}-k)!]^{1/2}},
\notag \\
B_{m_{1}k}^{(o)} &=&\frac{(\sqrt{2})^{2k+1}(\frac{(N+m_{1})!(N-m_{1})!}{2N!}%
)^{1/2}\sqrt{N!}}{[(2k+1)!(\frac{N-m_{1}-1}{2}-k)!(\frac{N+m_{1}-1}{2}%
-k)!]^{1/2}}. \hskip 18pt
\end{eqnarray}%
We calculate the particle numbers and number fluctuations in the AA phase
and find that the average numbers of atoms in the six components are exactly
all equal,
\begin{eqnarray}
\left\langle n_{1}^{(j)}\right\rangle =\left\langle n_{0}^{(j)}\right\rangle
=\left\langle n_{-1}^{(j)}\right\rangle =N/3,
\end{eqnarray}%
i.e., the condensate is fragmented \cite{HoYip}. The fluctuations are given
explicitly
\begin{eqnarray}
\left\langle \Delta n_{0}^{(j)}\right\rangle &=&\frac{\sqrt{N^{2}+9N}}{3%
\sqrt{5}},  \notag \\
\left\langle \Delta n_{\pm 1}^{(j)}\right\rangle &=&\frac{2\sqrt{N^{2}+3N/2}%
}{3\sqrt{5}},  \label{flu}
\end{eqnarray}%
which approximatively satisfy $\left\langle \Delta n_{1}^{(j)}\right\rangle
=2\left\langle \Delta n_{0}^{(j)}\right\rangle =\left\langle \Delta
n_{-1}^{(j)}\right\rangle $ for large $N$, as opposed to $2\left\langle
\Delta n_{1}\right\rangle =\left\langle \Delta n_{0}\right\rangle
=2\left\langle \Delta n_{-1}\right\rangle $ for the single species singlet
state ($c_{2}>0$) \cite{HoYip}. %We attribute this effect to the strong
%anti-ferromagnetic coupling between the ferromagnetic and polar species.

\begin{figure}[tbp]
\includegraphics[width=3.2in]{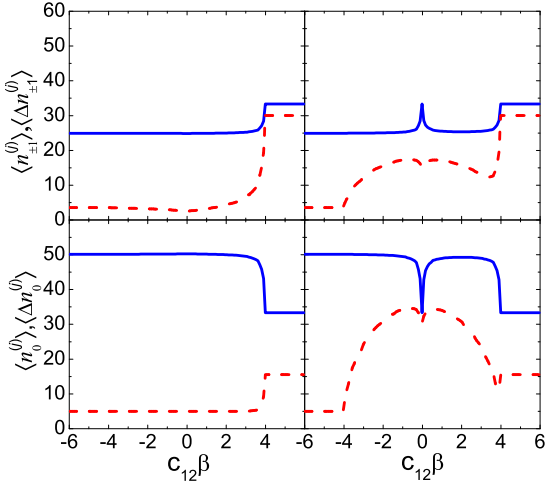}
\caption{(Color online) The dependence of atom numbers and fluctuations on $%
c_{12}\protect\beta $ at fixed values of $c_{1}\protect\beta _{1}=-1$ and $%
c_{2}\protect\beta _{2}=2$. Blue solid (red dashed) lines denote $%
\left\langle n_{\protect\alpha }^{(j)}\right\rangle $ ($\left\langle \Delta
n_{\protect\alpha }^{(j)}\right\rangle $). Note $\left\langle
n_{+1}^{(j)}\right\rangle =\left\langle n_{-1}^{(j)}\right\rangle $ and $%
\left\langle \Delta n_{+1}^{(j)}\right\rangle =\left\langle \Delta
n_{-1}^{(j)}\right\rangle $. Left (right) column denotes the ferromagnetic
(polar) condensate.}
\label{fig4}
\end{figure}

For a comprehensive understanding of the fluctuations in the entire
parameter region, we consider the direct product of the Fock states of the
two species $\left\vert n_{1}^{(1)},n_{0}^{(1)},n_{-1}^{(1)}\right\rangle
\otimes \left\vert n_{1}^{(2)},n_{0}^{(2)},n_{-1}^{(2)}\right\rangle $,
which may be equivalently defined as
\begin{eqnarray}
&&\hat{n}_{\alpha }^{(1,2)}\left\vert
n_{0}^{(1)},m_{1},n_{0}^{(2)},m_{2};m\right\rangle  \notag \\
&=&n_{\alpha }^{(1,2)}\left\vert
n_{0}^{(1)},m_{1},n_{0}^{(2)},m_{2};m\right\rangle .
\end{eqnarray}%
Here $m_{1}$ and $m_{2}$ are the corresponding magnetization specified as $%
m_{j}=n_{1}^{(j)}-n_{-1}^{(j)}$ and $m=m_{1}+m_{2}$ is the total
magnetization. For simplification, we restrict ourselves into the subspace
that the total magnetization is conserved $m=0$, in which case all states
are non-degenerate. Using the full quantum approach of exact
diagonalization, we simulate the distribution in this subspace with $%
N_{1}=N_{2}=100$, and illustrate the dependence of the particle numbers and
fluctuations for the six components on $c_{12}\beta $ in Fig. \ref{fig4}.

For very large values of $|c_{12}\beta |$, the distribution of atoms and
fluctuations behave uniformly, indicating typical
ferromagnetic(FF)/anti-ferromagnetic(AA) spin-exchange features. The number
distributions for all components are essentially the same over the entire
region, except for the case of $c_{12}\beta =0$. A tiny $c_{12}\beta \neq 0$
brings the system into the ferromagnetic-like distribution, i.e. $%
2\left\langle n_{1}^{(j)}\right\rangle =\left\langle
n_{0}^{(j)}\right\rangle =2\left\langle n_{-1}^{(j)}\right\rangle =N/2$,
consistent with the prediction that the ferromagnetic condensate is more
stable. The anti-ferromagnetic distributions for the six components are the
same $\left\langle n_{1}^{(j)}\right\rangle =\left\langle
n_{0}^{(j)}\right\rangle =\left\langle n_{-1}^{(j)}\right\rangle $, in
agreement with the analytical results (\ref{flu}). The fluctuations are, on
the other hand, quite different for positive or negative $c_{12}\beta $. In
the FF phase ($c_{12}\beta \,<-4$) the fluctuations for both species are
small ($\sim \sqrt{N}$). In the AA phase ($c_{12}\beta \,>4$) the
fluctuations are large and approximatively satisfy $\left\langle \Delta
n_{1}^{(j)}\right\rangle =2\left\langle \Delta n_{0}^{(j)}\right\rangle
=\left\langle \Delta n_{-1}^{(j)}\right\rangle $ (with $\left\langle \Delta
n_{\pm 1}^{(j)}\right\rangle \approx 30.03$ and $\left\langle \Delta
n_{0}^{(j)}\right\rangle \approx 15.56$). In the regions of MM$_{-}$ and MM$%
_{+}$ ($-4<c_{12}\beta \,<4$), polar atoms exhibit larger fluctuations than
the ferromagnetic ones, which change quadratically with $c_{12}\beta $. When
$c_{12}\beta \,>0,$ as the inter-species coupling increases, the fluctuation
of polar atoms dramatically changes from $2\left\langle \Delta
n_{1}^{(2)}\right\rangle =\left\langle \Delta n_{0}^{(2)}\right\rangle
=2\left\langle \Delta n_{-1}^{(2)}\right\rangle $ to $\left\langle \Delta
n_{1}^{(2)}\right\rangle =2\left\langle \Delta n_{0}^{(2)}\right\rangle
=\left\langle \Delta n_{-1}^{(2)}\right\rangle $ (approximatively).

Before concluding, we show the difference of the singlet state $\left\vert
N,N,0,0\right\rangle $ and the fully-paired state $Z^{1/2}(\hat{\Theta}%
_{12}^{\dag })^{N}\left\vert 0\right\rangle $ from the viewpoint of quantum
fluctuation \cite{Shi09,Xu2}. Using the method of generating function \cite%
{Xu2}, the difference between the two states can be shown by taking $%
N_{1}=N_{2}=2$ as an example. The result remains the same from the angular
momentum theory, i.e.
\begin{eqnarray}
\left\vert 2,2,0,0\right\rangle &=&\underset{m_{1},m_{2}}{\sum }%
C_{F_{1,}m_{1};F_{2,}m_{2}}^{F=0,m=0}\left\vert 2,m_{1}\right\rangle
\left\vert 2,m_{2}\right\rangle  \notag \\
&=&\frac{1}{2\sqrt{5}}(\hat{\Theta}_{12}^{\dag 2}-\frac{1}{3}\hat{A}^{\dag }%
\hat{B}^{\dag })\left\vert 0\right\rangle .
\end{eqnarray}%
More generally, according to the multinomial theorem
\begin{eqnarray}
\left( x_{1}+x_{2}+x_{3}\right)
^{n}=\sum_{k=0}^{n}\sum_{l=0}^{k}c_{nlk}x_{1}^{n-k}x_{2}^{k-l}x_{3}^{l},
\end{eqnarray}%
with $c_{nlk}=n!/\left( l!(k-l)!(n-k)!\right) $, we find that the state $(%
\hat{\Theta}_{12}^{\dag })^{N}\left\vert 0\right\rangle $ can be described
by the Fock state $\left\vert
n_{1}^{(1)},n_{0}^{(1)},n_{-1}^{(1)}\right\rangle \otimes \left\vert
n_{1}^{(2)},n_{0}^{(2)},n_{-1}^{(2)}\right\rangle $ as%
\begin{eqnarray}
&&(\hat{\Theta}_{12}^{\dag })^{N}\left\vert 0\right\rangle  \notag \\
&=&\left( \hat{a}_{0}^{\dag }\hat{b}_{0}^{\dag }-\hat{a}_{1}^{\dag }\hat{b}%
_{-1}^{\dag }-\hat{a}_{-1}^{\dag }\hat{b}_{1}^{\dag }\right) ^{N}\left\vert
0\right\rangle  \notag \\
&=&\sum_{k=0}^{N}\sum_{l=0}^{k}c_{Nlk}(\hat{a}_{0}^{\dag }\hat{b}_{0}^{\dag
})^{N-k}(-\hat{a}_{1}^{\dag }\hat{b}_{-1}^{\dag })^{k-l}(-\hat{a}_{-1}^{\dag
}\hat{b}_{1}^{\dag })^{l}\left\vert 0\right\rangle  \notag \\
&=&\sum_{k=0}^{N}\sum_{l=0}^{k}(-1)^{k}N!\left\vert k-l,N-k,l\right\rangle
\otimes \left\vert l,N-k,k-l\right\rangle, \hskip 18pt
\end{eqnarray}%
where we have used the property $\left( \hat{a}^{\dag }\right)
^{N}\left\vert 0\right\rangle =\sqrt{N!}\left\vert N\right\rangle $. We
calculate the atom numbers and fluctuations for the two states $\left\vert
N,N,0,0\right\rangle $ and $Z^{1/2}(\hat{\Theta}_{12}^{\dag })^{N}\left\vert
0\right\rangle $, and find that one cannot distinguish them from the atom
number distributions%
\begin{eqnarray}
\left\langle n_{1}^{(j)}\right\rangle =\left\langle n_{0}^{(j)}\right\rangle
=\left\langle n_{-1}^{(j)}\right\rangle =N/3,
\end{eqnarray}%
which are the same. The fluctuations, however, reveal the secret. For state $%
Z^{1/2}(\hat{\Theta}_{12}^{\dag })^{N}\left\vert 0\right\rangle $, the
number fluctuations are equally distributed, i.e.
\begin{eqnarray}
\left\langle \Delta n_{1}^{(j)}\right\rangle &=&\left\langle \Delta
n_{0}^{(j)}\right\rangle =\left\langle \Delta n_{-1}^{(j)}\right\rangle
\notag \\
&=&\sqrt{N(N+1)/6-N^{2}/9},
\end{eqnarray}%
different from the results for the state $\left\vert N,N,0,0\right\rangle $
we obtained in Eq. (\ref{flu}).

\section{Conclusion}

To conclude, we study atom number distributions and fluctuations for the
ground state of a mixture of two spin-1 condensates, one being ferromagnetic
and the other being polar, in the absence of a $B$-field. For all possible
inter-species coupling parameter $c_{12}\beta $, the exact quantum states
are constructed from angular momentum algebra which are further expanded
into the six-component Fock states of the mixture. The ground state quantum
phases are classified into four types according to $c_{12}\beta $. The most
interesting AA phase of a singlet for the total spin, where spins of each
species are polarized in opposite directions, is fragmented with six
components equally distributed $\left\langle n_{1}^{(j)}\right\rangle
=\left\langle n_{0}^{(j)}\right\rangle =\left\langle
n_{-1}^{(j)}\right\rangle =N/3$ and exhibits anomalous number fluctuations.
We find that the fluctuations satisfy $\left\langle \Delta
n_{1}^{(j)}\right\rangle =2\left\langle \Delta n_{0}^{(j)}\right\rangle
=\left\langle \Delta n_{-1}^{(j)}\right\rangle $ for large $N$, different
from the result of single species polar state ($c_{2}>0$) $2\left\langle
\Delta n_{1}\right\rangle =\left\langle \Delta n_{0}\right\rangle
=2\left\langle \Delta n_{-1}\right\rangle $. Our results highlight the
significant promises for experimental work on Na and Rb atomic condensate
mixtures since optical Feshbach resonances make it possible to tune the spin
exchange interaction.

This work is supported by the NSF of China under Grant No. 10774095, the NSF
of Shanxi Province under Grant No. 2009011002, the National Basic Research
Program of China (973 Program) under Grant Nos. 2006CB921102, 2010CB923103,
2006CB921206 and 2006AA06Z104, and the Program for New Century Excellent
Talents in University (NCET).

\end{document}